# Backflow-Induced Inertial Arrest of Velocity Fluctuations in Sedimenting Suspensions

Hsien-Hung Wei*

*Department of Chemical Engineering, National Cheng Kung University, Tainan 701, Taiwan*

A self-contained hydrodynamic theory is proposed to reconcile the discrepancy between divergent Stokesian velocity fluctuations and finite experimental measurements in sedimenting suspensions. We show that the compensating backflow induces non-negligible inertia, giving rise to an emergent screening length $\xi \sim a\ \phi^{-1/3} Re_p^{-1/3}$ far exceeding the mean interparticle spacing $a\phi^{-1/3}$ at vanishingly small particle Reynolds numbers. This backflow inertial screening, together with finite-time viscous diffusion, arrests the indefinite spatiotemporal growth of large-scale velocity fluctuations. The resulting velocity fluctuations scale as $\delta u \sim \phi^{1/3} V_s Re_p^{-1/6}$, together with the viscous correlation time $\tau_c=\xi^2/\nu$, reproducing the well-known hydrodynamic self diffusivity scaling $D_H \sim V_s a$. The theory predicts the prefactors of these scaling laws without adjustable parameters, in good quantitative agreement with experimental measurements. It also successfully captures the experimentally observed crossover from the finite-correlation regime to the finite-system regime as the screening length becomes comparable to the system size.

In low Reynolds number hydrodynamics, the long-range nature of hydrodynamic interactions in sedimenting suspensions is known to give rise to a remarkable divergence of velocity fluctuations in the thermodynamic limit. This so-called Caflisch-Luke (CL) paradox originates from the slow $1/r$ decay of the Stokeslet-induced velocity field generated by each settling particle. Under the assumptions of steady, inertialess Stokes flow, the CL theory as well as the heuristic Poisson fluctuation argument of Hinch [2] predicted that the velocity variance $\langle u'^2\rangle$ of a homogeneous suspension increases without bound with system size $L$ according to

$$\langle u'^2\rangle \sim \phi V_s^2 (L/a), \qquad (1)$$

where $V_s$ is the Stokes terminal velocity of an isolated falling particle of radius $a$, and $\phi$ is the particle volume fraction.

However, experiments consistently reported finite velocity fluctuations, with no evidence of unbounded growth with system size. In particular, Segrè *et al.* [3] reported an unusually large correlation length $\xi \sim 20\ a\phi^{-1/3}$ that greatly exceeds the mean interparticle spacing $a\phi^{-1/3}$ and velocity fluctuations $\delta u \sim 2\phi^{1/3} V_s$. These features were also observed in systematic measurements by Guazzelli and co-workers [4-6]. Moreover, the measured hydrodynamic self diffusivity is $D_H \sim V_s a$ [7,8], implying that velocity fluctuations are also correlated over a finite characteristic time rather than over an indefinitely growing time scale. A comprehensive review by Guazzelli and Hinch [9] further highlighted the remarkable robustness of these scaling behaviors across different experiments.

On the theoretical side, a variety of mechanisms have been proposed to either introduce effective cutoffs or suppress velocity fluctuations at large scales. They include deficit-induced hydrodynamic screening [10], inertial/Oseen screening [2,11], sidedwall restraining [12], density stratification [13-15], nonlinear stochastic coupling [16], and hyperuniform ordering [17].

Despite these advances, a general theory that simultaneously accounts for finite spatial and temporal correlation behaviors in quantitative agreement with experimental observations remains lacking.

In this work, we develop such a theory. Our approach builds upon three key developments in the sedimentation literature: (i) Batchelor's identification of the compensating backflow necessary for cancelling unbounded collective Stokeslet disturbances [18], (ii) Hinch's recognition of the potential role of inertia in providing a large-scale cutoff [2], and (iii) Guazzelli and co-workers' falling-cloud picture [19,20] together with their recent experimental characterization of weak inertial effects on velocity fluctuations [4]. We show that these ideas can be unified into a self-consistent hydrodynamic mechanism—inertia associated with the compensating backflow can arrest the indefinite growth of large-scale velocity fluctuations. This mechanism provides a natural regularization of the CL divergence without invoking an ad hoc cutoff.

*Correlation scalings*— We begin with the experimental observation [3-5] that an initially large-scale stirring fluid motion gradually decay and reorganize over time into smaller-scale fluctuating vortices as the suspension evolves toward a steady state. This observation implies that there may be damping effects on large-scale velocity fluctuations, turning them into a steady coherent swirl structure.

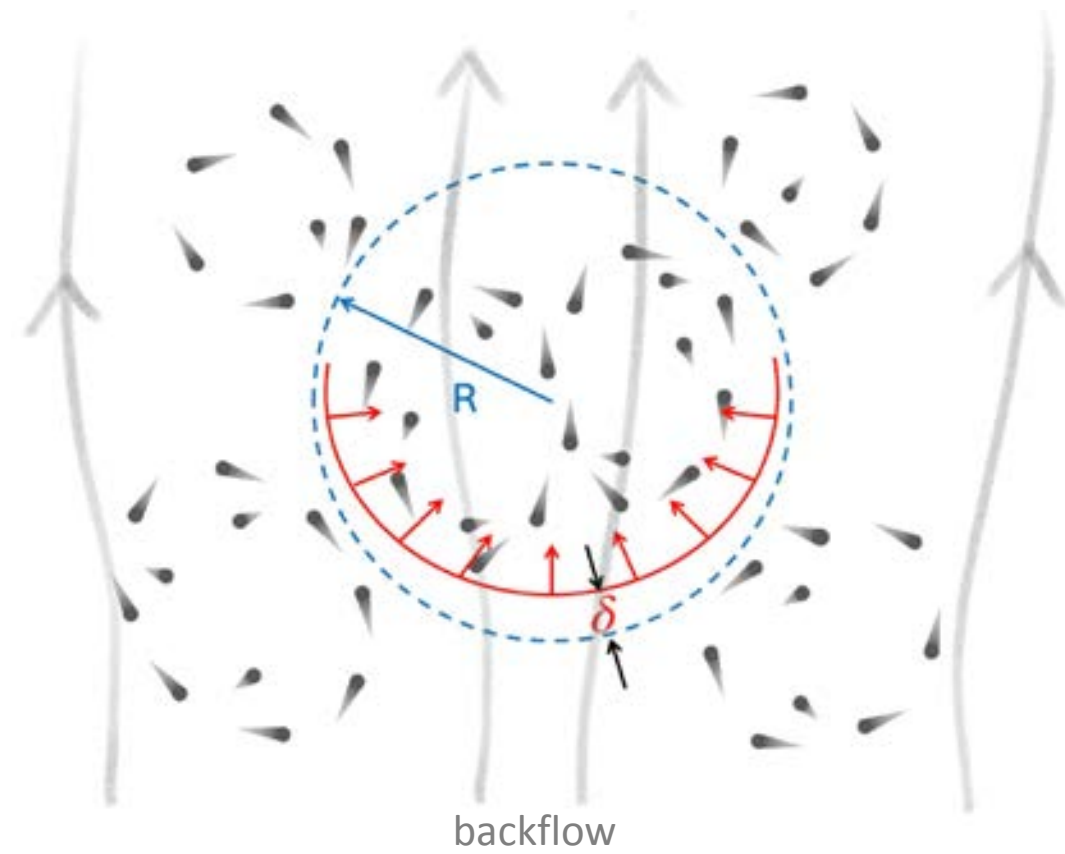

FIG.1. Schematic illustration of the cloud-backflow model, showing the formulation and evolution of the correlation boundary layer in a sedmenting suspension.

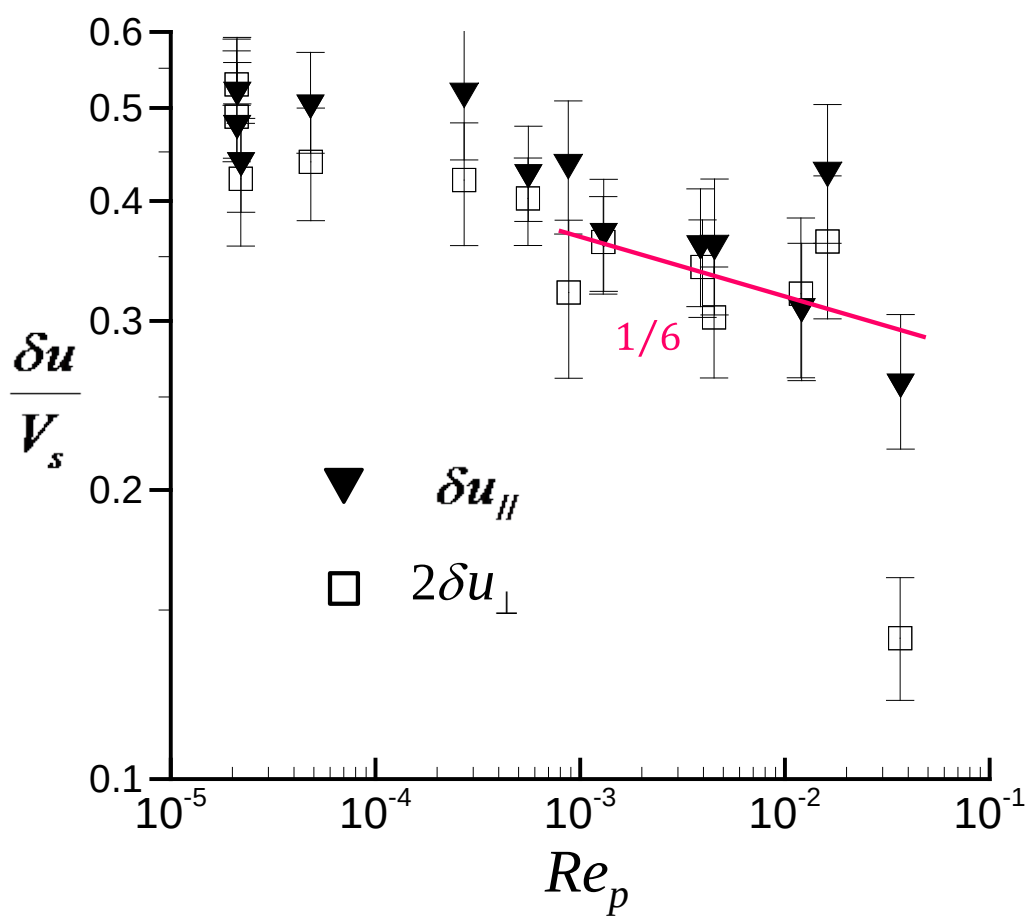

FIG.2. Velocity fluctuations measured by Bergougnoux and Guazzelli [4], exhibiting a $Re_p^{-1/6}$ dependence well captured by the scaling relation (3). This weak decrease with $Re_p$ manifests beyond the plateau region at small but sufficiently large $Re_p$, supporting the proposed inertial suppression mechanism.

As illustrated in FIG. 1, the coherent swirls can be viewed as interacting falling clouds. We consider a test cloud of radius $R$, analogous to a cloud settling in vortices [19] but immersed instead in the compensating upward backflow generated by the surrounding suspension. The cloud settles collectively at a velocity $U_{cloud} \sim N\, V_s a/R$ (sum of Stokeslets) $\sim c_0\; V_s a\; R^2$ [20], where $N \sim c_0\; R^3$ is the number of particles within the cloud and $c_0 \sim \phi/a^3$ is the particle concentration. Owing to shearing by the opposing backflow, a layer of concentrated vorticity ($\varpi$) is preferentially generated at the lower cloud boundary. This vorticity layer, of thickness $\delta$, subsequently evolves inward via viscous diffusion of scale $\nu\varpi/\delta^2$, thus forming the correlated boundary layer.

In the meantime, these vorticity disturbances are advected by the compensating backflow $U_{back} \sim U_{cloud}$, which carries inertial momentum into the cloud interior in the form of a vorticity flux $\sim U_{back}\; \varpi$. At this point, the inertial advection of vorticity over the cloud scale $U_{back}\,\varpi/R$ becomes comparable to the viscous diffusion of vorticity $\nu\varpi/R^2$, where $\nu$ is the fluid kiematic viscosity. Equating these two rates yields the cloud Reynolds number $Re_{cloud} = U_{back}R/\nu \sim c_0\; R^3\; Re_p = O(1)$ in terms of the particle Reynolds number $Re_p = V_s a/\nu$ (with ν being the kinematic viscosity). This condition provides a cutoff radius $R_{cutoff} \;\sim a\;(\phi\; Re_p)^{-1/3}$ that limits the spatial growth of velocity disturbances. Beyond this cutoff, viscous diffusion of vorticity outpaces the advective supply from the backflow, preventing the boundary layer from ever fully engulfing the cloud core. Consequently, $R_{cutoff}$ defines a characteristic hydrodynamic correlation length

$$\xi \sim a\; \phi^{-1/3}\; Re_p^{-1/3}, \tag{2}$$

which diverges as $Re_p \rightarrow 0$. Thus, even a vanishingly small $Re_p$ can produce $\xi$ far exceeding the mean particle spacing $a\; \phi^{-1/3}$. This naturally explains $\xi \approx 20\; a\; \phi^{-1/3}$ reported by Segré *et al.* [3], since (2) yields $\xi \sim 26\; a\; \phi^{-1/3}$ with $Re_p \approx 0.6\times10^{-4}$ in their work. Those measured in the recent experiments [4] are likewise of the same order of magnitude and can be interpreted in the same way.

The resulting velocity fluctuations can be estimated using the CL scaling (1) with $L \sim \xi$ and becomes finite:

$$\delta u \sim \phi^{1/3}\; V_s\; Re_p^{-1/6}, \tag{3}$$

which becomes unbounded as $Re_p \rightarrow 0$. With $Re_p \approx 0.6\cdot10^{-4}$ in Segré *et al*'s work [3], the predicted $\delta u \sim 5.1\phi^{1/3}V_s$ is also close to the measured value $2.0\phi^{1/3}V_s$. FIG. 2 shows that the predicted $Re_p^{-1/6}$ dependence agrees with the trend observed in the recent experiments of Bergougnoux and Guazzelli [4], supporting the inertial suppression mechanism [21].

As velocity fluctuations can only propagate through vorticity diffusion across the correlated region, the correlation time is the corresponding viscous diffusion time $\tau_c = \xi^2/\nu$, yielding

$$\tau_c \sim (a^2/\nu)\; \phi^{-2/3}\; Re_p^{-2/3}, \tag{4}$$

which diverges as $Re_p \rightarrow 0$. So a finite but small $Re_p$ yields a finite $\tau_c$ much longer than the particle viscous damping time $a^2/\nu$. The resulting hydrodynamic self diffusivity $D_H \sim (\delta u)^2\; \tau_c$ is also finite, recovering the well-known scaling

$$D_H \;\sim V_s\, a, \tag{5}$$

which is independent of both $Re_p$ and $\phi$ as reported in experiments [7,8].

*Backflow inertia screening*– the scaling results presented above are derived from the heuristic cloud-backflow picture, from which the finite correlation length given by (2) is inferred. We now show that the same characteristic correlation length emerges naturally from the Oseen-like inertial screening in the presence of the backflow.

We start with the incompressible Navior-Stokes equations for velocity $\boldsymbol{u}$ and pressure $p$ in the suspension driven by sum of point forces exerted by settling particles:

$$\nabla\cdot \boldsymbol{u} = 0. \tag{6a}$$

$$\rho(\boldsymbol{u}_t + \boldsymbol{u}\cdot\nabla\boldsymbol{u}) = -\nabla p + \eta\nabla^2\boldsymbol{u} + \sum\delta(\mathbf{x}-\mathbf{x}_n)\boldsymbol{F}^{(n)}, \tag{6b}$$

where $\mathbf{x}_n$ denotes the position of the $n$-th particle and $\boldsymbol{F}^{(n)}$ is the corresponding point force. $\rho$ and $\eta$ denote the fluid's density and viscosity, respectively. In the Stokes limit, these particle forces generate the Stokeslet collective flow field:

$$\boldsymbol{u}(\mathbf{x}) = (8\pi\eta)^{-1}\sum \mathbf{G}(\mathbf{x}-\mathbf{x}_n)\cdot \boldsymbol{F}^{(n)}, \quad (7)$$

where $\mathbf{G}(\mathbf{x}) = \mathbf{I}/r + \mathbf{xx}/r^3$ is the Oseen tensor with $r=|\mathbf{x}|$.

The particle motion generates particle density fluctuations $c'$, which in turn induce fluid disturbance velocity $\boldsymbol{u}'$ and pressure $p'$. According to Saffman [22], these disturbance quantities are physically observable only after removing the background $k$=0 Fourier mode from (7), corresponding to the infinite-system limit $L\to\infty$. Physically, this is equivalent to examining the response with respect to the zero-mean-flow condition $\langle \boldsymbol{u}\rangle_0=0$ required for sedimenting suspensions [18].

To enforce $\langle \boldsymbol{u}\rangle_0=0$ globally through incorporation of such background mode, a physically observable $\boldsymbol{u}'$, following the same spirit as Saffman [22], can be constructed as

$$\boldsymbol{u}' = \boldsymbol{u} - \boldsymbol{u}_{back}, \quad (8)$$

by subtracting the compensating backflow, i.e., the ensemble averaged collective Stokeslet field for $|\mathbf{x}| < L$:

$$\boldsymbol{u}_{back}(\mathbf{x}) = U_\infty(L) + \boldsymbol{U}_{back}(\mathbf{x}), \quad (9)$$

based on the analogous Hill's spherical vortex solution of a uniformly settling particle cloud [19]. The first term $U_\infty(L) = \langle \boldsymbol{f}\rangle L^2/3\eta$ is spatially uniform and exactly cancels the mean of the collective Stokeslet field in (7) as $L\to\infty$, thereby recovering Batchelor's renormalization [18]. Here $\langle \boldsymbol{f}\rangle = c_0\langle \boldsymbol{F}\rangle$ is the mean Stokes forcing of $\langle \boldsymbol{F}\rangle = 6\eta V_s a\ \mathbf{e}$ acting in the direction of gravity $\mathbf{e}$. $\boldsymbol{U}_{back}(\mathbf{x})$ is the leading-order spatially varying (quadratic) component of the backflow:

$$\boldsymbol{U}_{back}(\mathbf{x}) = (15\eta)^{-1}[\langle \boldsymbol{f}\rangle\cdot\mathbf{xx} - 2|\mathbf{x}|^2\langle \boldsymbol{f}\rangle], \quad (10)$$

accounting for the macroscopic flow variation within $|\mathbf{x}| << L$, sufficiently away from the system's boundaries. Moreover, the corresponding vorticity field $\nabla\times\boldsymbol{U}_{back}(\mathbf{x}) = \langle \boldsymbol{f}\rangle/3\eta \times \mathbf{x}$ increases linearly with $\mathbf{x}$, endowing the backflow with an intrinsic large-scale rotational structure. This feature may provide a natural basis for the emergence of the large-scale swirling motions observed in experiments [5].

Transforming to the reference frame moving with $U_\infty(L)$, the remaining quadratic backflow $\boldsymbol{U}_{back}(\mathbf{x})$ serves as the macroscopic background flow. The physically observable disturbance variables $\boldsymbol{u}' = \boldsymbol{u} - \boldsymbol{U}_{back}$, $p' = p - p^*_{back}$ and $c' = c - c_0$ are then governed, to leading-order, by substituting these decompositions into (6) while neglecting the nonlinear term $\boldsymbol{u}'\cdot\nabla\boldsymbol{u}'$ [21]:

$$\nabla\cdot\boldsymbol{u}' = 0, \quad (11a)$$

$$\rho(\boldsymbol{u}'_t + \nabla\cdot(\boldsymbol{U}_{back}\boldsymbol{u}' + \boldsymbol{u}'\boldsymbol{U}_{back}) = -\nabla p' + \eta\nabla^2\boldsymbol{u}' + c'\langle \boldsymbol{F}\rangle, \quad (11b)$$

where $p^*_{back} = p_{back} - \rho\Phi$ absorbs the backflow inertial potential $\Phi = |\langle \boldsymbol{f}\rangle|^2(|\mathbf{x}|^2(\mathbf{e}\cdot\mathbf{x})^2 - (1/2)|\mathbf{x}|^4)/225\eta^2$ from $\nabla\Phi = \boldsymbol{U}_{back}\cdot\nabla\boldsymbol{U}_{back}$ because $\nabla\times(\boldsymbol{U}_{back}\cdot\nabla\boldsymbol{U}_{back}) = 0$.

Next we perform Fourier transform for (11) with

$$(\hat{\boldsymbol{u}}', \hat{p}', \hat{c}')(\mathbf{k},\omega) = \int_{-\infty}^{\infty}\int_{-\infty}^{\infty}(\boldsymbol{u}', p', c')(\mathbf{x},t)e^{-i\mathbf{k}\cdot\mathbf{x}+i\omega t}d\mathbf{x}dt. \quad (12)$$

Applying the projection tensor $\mathbf{P} = \mathbf{I} - \mathbf{kk}/k^2$ to eliminate the pressure term with $\mathbf{P}\cdot\mathbf{k}\,\hat{p}'=0$ and write $\mathbf{P}\cdot\hat{\boldsymbol{u}}' = \hat{\boldsymbol{u}}'$ using the incompressibility $\mathbf{k}\cdot\hat{\boldsymbol{u}}'=0$ from (11a), we obtain

$$(-i\omega/\nu + k^2)\hat{\boldsymbol{u}}' + (2\pi/5)c_0 Re_p i\,\mathbf{P}\cdot(\mathbf{k}\cdot\mathbf{T}[\hat{\boldsymbol{u}}']) = \hat{c}'\,\mathbf{P}\cdot\langle \mathbf{F}\rangle/\eta. \quad (13)$$

Here $\mathbf{T}[\hat{\boldsymbol{u}}'] = -\mathbf{e}\cdot\nabla_k[\nabla_k\hat{\boldsymbol{u}}' + (\nabla_k\hat{\boldsymbol{u}}')^T] + 2[\mathbf{e}\nabla_k^2\hat{\boldsymbol{u}}' + (\mathbf{e}\nabla_k^2\hat{\boldsymbol{u}}')^T]$ is the Fourier transform for the backflow inertial term in (11b), in which $\nabla_k\nabla_k$ and $\nabla_k^2$ arise from Fourier transforms $-(2\pi)^3\nabla_k\nabla_k\,\delta(\mathbf{k})$ for $\mathbf{xx}$ and $-(2\pi)^3\nabla_k^2\delta(\mathbf{k})$ for $|\mathbf{x}|^2$ in (10). These differential operators are precisely what allow the emergence of a screening length through balancing the $\mathbf{k}\cdot\mathbf{T}[\hat{\boldsymbol{u}}']$ term to the viscous $k^2$ term.

We use the local isotropic approximation to elucidate such emergence. This approximation may be reasonable, as experimentally observed vigorous mixing within correlated swirl structures suggests a tendency toward local isotropy of the velocity fluctuations [3-5]. By adopting $\nabla_k \approx (\mathbf{k}/k)\mathrm{d}/\mathrm{d}k$ and using identities $\mathbf{k}\cdot\mathrm{d}\hat{\boldsymbol{u}}'/\mathrm{d}k=0$ and $\mathbf{k}\cdot\mathrm{d}^2\hat{\boldsymbol{u}}'/\mathrm{d}k^2=0$ from differentiation of $\mathbf{k}\cdot\hat{\boldsymbol{u}}'=0$, (13) reduces to

$$[-i\omega/\nu + k^2 + A\,c_0 Re_p\,i\,(k\,\mathrm{d}^2/\mathrm{d}k^2 + 4\mathrm{d}/\mathrm{d}k)]\hat{\boldsymbol{u}}' = \hat{c}'\,\mathbf{P}\cdot\langle \mathbf{F}\rangle/\eta. \quad (14)$$

Here $(\mathbf{e}\cdot\mathbf{k})$ in $\mathbf{k}\cdot\mathbf{T}[\hat{\boldsymbol{u}}']$ has been approximated by its root-mean-square value $\langle \mathbf{e}\cdot\mathbf{k}\rangle_{rms} \approx k/\sqrt{3}$ through angular average. $A = 2\pi\sqrt{3}/15$ is the geometric factor.

As indicated by (14), in the small $k$ regime the backflow inertial term dominates with a boundary-layer-like structure $\sim Ac_0Re_p/k$, whereas the viscous dissipation $k^2$ term prevails for large $k$. Balancing these two contributions yields the characteristic crossover

$$k^* = A'^{1/3}a^{-1}(\phi Re_p)^{1/3}, \quad (15)$$

with $A' = 3A/4\pi = \sqrt{3}/10$ The associated characteristic length scale $1/k^*$ ($>>a$) is precisely the screening length given by (2). Dynamically, the transition from inertial propagation at small $k$ to viscous dissipation at large $k$ occurs at $\omega^* = \nu\,k^{*2}$, yielding

$$\omega^* = (\nu/a^2)\,A'^{2/3}(\phi Re_p)^{2/3}, \quad (16)$$

whose inverse is exactly the correlation time given by (4). This frequency (time) scale is much smaller (larger) than the particle viscous damping frequency $\nu/a^2$ (time $a^2/\nu$) because $\phi Re_p << 1$.

Such backflow inertial screening mechanism becomes particularly transparent in the large $k$ regime. In this regime, the WKB analysis identifies that the dominant terms in (14) are $k^2\hat{\boldsymbol{u}}' + i\,k^{*3}k\mathrm{d}^2\hat{\boldsymbol{u}}'/\mathrm{d}k^2$, yielding an asymptotic solution

$$\hat{\boldsymbol{u}}' = C\,\mathrm{Ai}(e^{i\pi/6}k/k^*) + \hat{c}'\,\mathbf{P}\cdot\langle \mathbf{F}\rangle/\eta k^2, \quad (17)$$

where $C$ is a constant. The Airy branch originates solely from the backflow inertia and provides a direct signature of the screening mechanism. Its Fourier inverse gives $(2\pi)^{-1}e^{-i\pi/6}k^*\exp(-k^{*3}r^3/3)$, demonstrating rapid attenuation over the distance $r \geq k^{*-1}$. Thus, $k^{*-1}$ naturally emerges as the characteristic spatial correlation length. The same localization is reflected in Fourier space: for $k >> k^*$, the asymptotic form $\mathrm{Ai}(z) \sim (2\sqrt{\pi})^{-1}z^{-1/4}\exp(-(2/3)z^{3/2})$ with $z = e^{i\pi/6}k/k^*$ shows exponential suppression of short wavelength modes. These real- and Fourier-space descriptions are therefore consistent and represent complementary manifestations of the same backflow-induced inertial screening. The remaining Stokeslet response $\hat{c}'\,\mathbf{P}\cdot\langle\mathbf{F}\rangle/\eta k^2$ is bounded at the crossover by $\hat{c}'(\mathbf{I}-\mathbf{ee})\cdot\langle\mathbf{F}\rangle/\eta k^{*2}$, limiting the amplitude of the residual small-scale velocity fluctuations.

In the small $k$ regime ($k \ll k^*$), corresponding to length scales larger than the screening length $k^{*-1}$, the asymptotic solution can be found as

$$\hat{\boldsymbol{u}}' = \hat{\boldsymbol{u}}'(0,\omega) + a_2 k^2 + O(k^3). \qquad (18)$$

The leading term $\hat{\boldsymbol{u}}'(0,\omega) = -i\,\hat{c}'(0,\omega)\langle\mathbf{F}\rangle\cdot(\mathbf{I}-\mathbf{ee})\rho/\omega$ is finite for non-zero $\omega$. If $\hat{c}'$ is assumed to undergo effective diffusive relaxation driven by conservative Gaussian noise, with the noise amplitude proportional to $k$ as a consequence of vigorous mixing within the correlated swirls, then $\hat{\boldsymbol{u}}'(0,\omega) \propto \hat{c}'(0,\omega) = 0$. The first surviving contribution is $O(k^2)$, arising from inertial-viscous suppression by backflow, and has magnitude $a_2 = -i\,\hat{c}'_1(0,\omega)\langle\mathbf{F}\rangle\cdot(\mathbf{I}-\mathbf{ee})/4\eta k^{*3}$, where $\hat{c}'_1$ is the expansion of $\hat{c}'$ at $O(k)$.

It should be noted that the small-$k$ expansion above is singular with respect to $Re_p$. As $Re_p \to 0$, $k^*$ vanishes and the screening length $\xi = k^{*-1}$ diverges according to (15), causing (18) to break down. In this limit, the backflow regularization mechanism disappears and the classical unscreened CL response is recovered, as indicated by (14). Thus, for a finite but small $Re_p$, $\xi$ may become comparable to or exceed the system size $L$. To preserve an interior screening region unaffected by the system's boundaries, it is necessary $\xi/L<1$. Using (15), this condition requires the particle inertia to be sufficiently strong, providing the screening criterion when $Re_p$ exceeds the threshold $Re_{p,\mathrm{TH}}$ according to

$$Re_p > Re_{p,\mathrm{TH}} \equiv A'^{-1}\phi^{-1}(L/a)^{-3}. \qquad (19)$$

For the experiments by Bergougnoux and Guazzelli [4] with $\phi$=0.3% and $L/a\approx 260$, (19) predicts $Re_{p,\mathrm{TH}} \approx 10^{-4}$, in good agreement with the observed value $4\times10^{-4}$. The corresponding system's Reynolds number $Re_{L,\mathrm{TH}} = Re_{p,\mathrm{TH}}(L/a) \approx 0.03$ is also close to the measured value 0.1. Below this threshold, $\xi/L>1$, so that the backflow screening mechanism cannot fully develop. The velocity fluctuations therefore remain bounded by the finite-system CL scaling given by (1), which explains the observed plateau, as shown in FIG.2.

Alternatively, the screening criterion (19) can be expressed in terms of the system size $L$ with respect to the mean interparticle spacing $a\phi^{-1/3}$:

$$L/a\phi^{-1/3} > \xi/a\phi^{-1/3} = A'^{-1/3}\,Re_p^{-1/3}. \qquad (20)$$

The critical value $\xi/a\phi^{-1/3}$ is precisely the screening length that varies as $Re_p^{-1/3}$ described by (15), controlling how the velocity fluctuation amplitude $\delta u$ varies with $L$. Below this value, unscreened $\delta u$ grows linearly with $L$ according to the CL scaling (1). Once $L$ reaches $\xi$ and beyond, the backflow screening becomes fully established and $\delta u$ becomes independent of $L$. This crossover explains the experimental observation by Segré *et al.* [3]. Also in their work, $Re_p\approx 0.6\times10^{-4}$ gives $\xi/a\phi^{-1/3}\approx 46$, reasonably capturing the observed value 150 in their experiments.

*Velocity fluctuations*–To quantify velocity fluctuations in the present spatiotemporal description, it is natural to compute the variance of the time-averaged velocity (denoted by an overbar) and express it in terms of the ensemble-averaged squared Fourier amplitude:

$$\overline{\left\langle\left\langle\left|\boldsymbol{u}'(\mathbf{x},t)\right|^2\right\rangle\right\rangle_V} = \frac{1}{VT}\int_{-\infty}^{\infty}\frac{d\mathbf{k}}{(2\pi)^3}\int_{-\infty}^{\infty}\frac{d\omega}{(2\pi)}\left\langle\left|\hat{\boldsymbol{u}}'(\mathbf{k},\omega)\right|^2\right\rangle, \qquad (21)$$

where $V$ is the system volume and $T$ is the total observation time. In deriving (21), we assume that the suspension is statistically homogeneous and ergodic. The connection between the variance and the squared Fourier amplitude follows directly from Parseval's theorem applied in both space and time.

Eq. (18) has established that in the small $k$ regime at $k \ll k^*$, $\hat{\boldsymbol{u}}'(k,\omega)$ is $O(k^2)$ due to backflow suppression. Also given that for $k \gg k^*$ the Airy branch is exponentially attenuated, leaving the bounded Stokeslet response as the dominant contribution according to (17), the variance in (21) may therefore be evaluated using the regularized Stokeslet solution $\hat{\boldsymbol{u}}' = \hat{G}(k,\omega)\,\hat{c}'(k)\,\langle\mathbf{F}\rangle\cdot(\mathbf{I}-\mathbf{ee})/\eta$ with $k \geq k^*$, where $\hat{G}(k,\omega) = (-i\omega/\nu + k^2)^{-1}$ is the Green function in free space. Therefore, the velocity variance, hereafter referred to as fluctuation intensity $(\delta u)^2$, can then be obtained as

$$\left(\delta u_\sigma\right)^2 = c_0\left(\frac{\left|\langle\mathbf{F}\rangle\right|}{\eta}\right)^2\int_{k^*}^{\infty}\frac{k^2 dk}{(2\pi)^3}S(k)\int\chi_\sigma^2 d\Omega\int_{-\infty}^{\infty}\frac{d\omega}{(2\pi)}\left|\hat{G}(k,\omega)\right|^2\mathcal{L}(k,\omega). \quad (22)$$

Here $\sigma$ denotes // or $\perp$ direction. $\chi_{//}=k_{//}^2/k^2$ and $\chi_\perp = -k_\perp k_{//}/k^2$ are the anisotropy factors. We have also invoked the density fluctuation spectrum $\langle|\hat{c}'(\mathbf{k},\omega)|^2\rangle = c_0 VTS(k)\mathcal{L}(k,\omega)$ in terms of the static structure factor $S(k)$ with finite $S(0)$ together with the Lorentzian diffusive relaxation function $\mathcal{L}(k,\omega) = 2D_{\mathrm{eff}}\,k^2/(\omega^2 + D_{\mathrm{eff}}^2 k^4)$, where $D_{\mathrm{eff}}$ is the effective particle diffusivity [23].

Performing the $\omega$ integration in (22) gives $k^{-4}(1+D_{\mathrm{eff}}/\nu)^{-1} \approx k^{-4}$, since $D_{\mathrm{eff}}/\nu$ is typically small [24]. The subsequent $k$-integration with the infrared cutoff $k^*$ yields $k^{*-1}S(0)/(2\pi)^3$, leading to the finite CL scaling $(\delta u)^2 \sim \phi V_s^2 (ak^*)^{-1}$ with the effective system size replaced by $k^{*-1}$. Since this result is dominated by $k^*$, it is insensitive to the detailed form of $S(k)$ as long as $S(k\to 0)$ remains finite. Therefore, in the screened regime $Re_p > Re_{p,\mathrm{TH}}$ with the screening length $\xi = 1/k^* < L$ in (20), the velocity fluctuation amplitude $\delta u \equiv \delta u_{//} = 2\delta u_\perp$, after including the anisotropy factors, can be evaluated as

$$\frac{\delta u}{\phi^{1/3}V_s} = \alpha'\left[S(0)\right]^{1/2}\left(\frac{\xi}{a\phi^{-1/3}}\right)^{1/2} = \alpha\ \left[S(0)\right]^{1/2} Re_p^{-1/6}, \qquad (23)$$

confirming the scaling relation (3), where $\alpha'=(36/5\pi)^{1/2}\approx 1.51$ and $\alpha=\alpha' A'^{-1/6}\approx 2.03$. The $Re_p^{-1/6}$ dependence also well captures the trend reported by Bergougnoux and Guazzelli [4], as shown in FIG. 2.

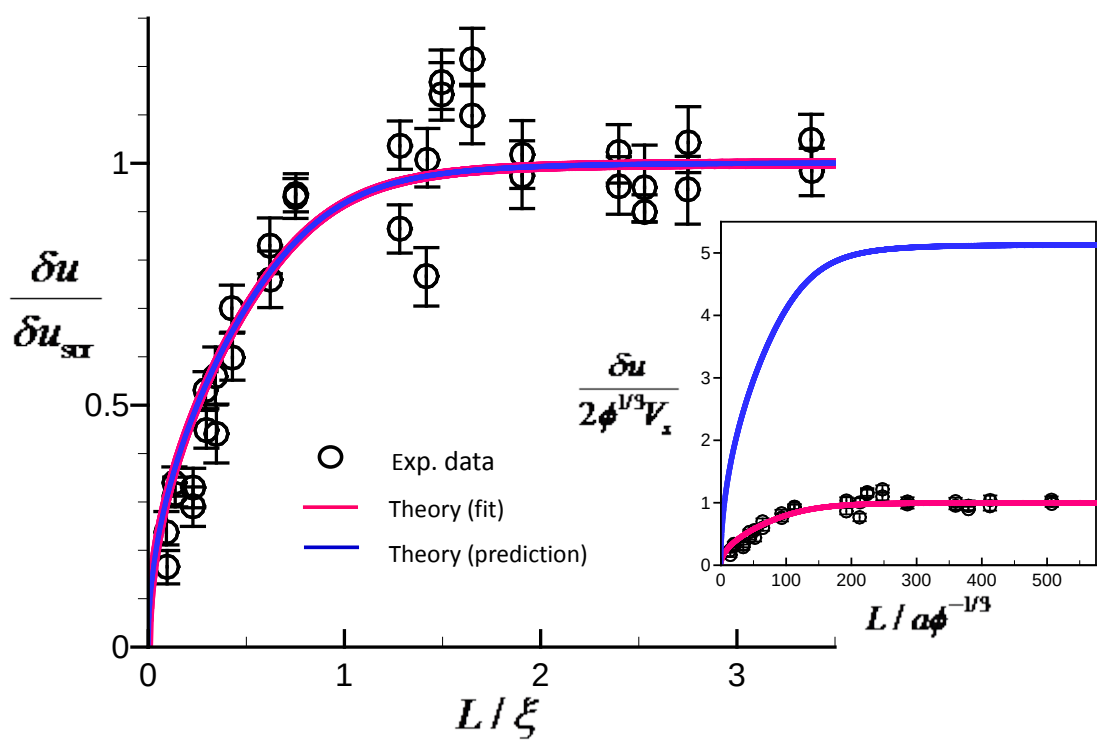

FIG. 3. Comparison of the experimental data (symbols) of Segré *et al*. [3] with the theoretical fit (pink lines) and the theoretical prediction (blue lines) based on Eq. (25).

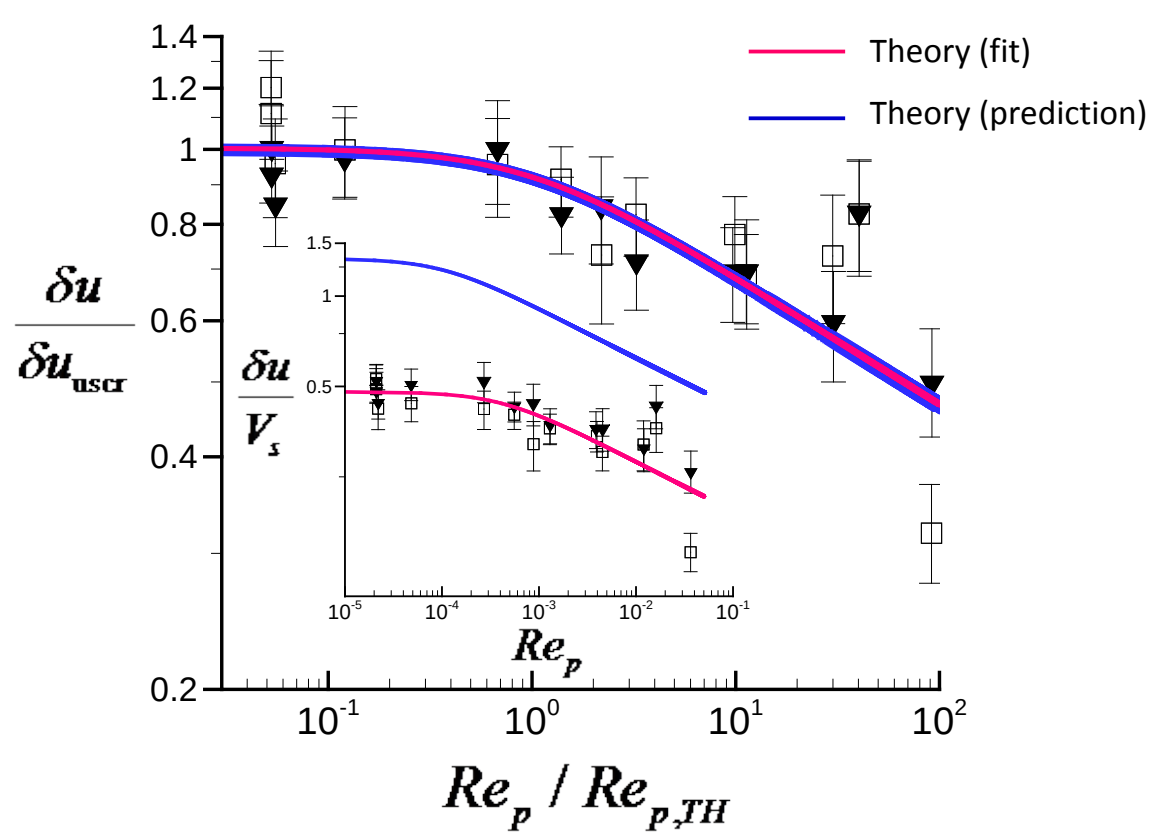

FIG. 4. Comparison of the experimental data (symbols) of FIG.2 with the theoretical fit (pink lines) and the theoretical prediction (blue lines) based on Eq. (26).

In the unscreened regime $Re_p < Re_{p,\mathrm{TH}}$ with $\xi > L$ violating (20), the screening length $\xi$ is limited by the system size $L$. Hence, $\delta u$ described by (23) has to saturate at $\xi$=$L$ or $Re_p$=$Re_{p,\mathrm{TH}}$ from (19), giving the finite-system CL result $\delta u = \alpha' [S(0)]^{1/2} \phi^{1/2} V_s (L/a)^{1/2}$. In the re-scaled form, it reads

$$\frac{\delta u}{\phi^{1/3}V_s} = \alpha'\left[S(0)\right]^{1/2}\left(\frac{L}{a\phi^{-1/3}}\right)^{1/2}. \qquad (24)$$

The screened velocity fluctuations $\delta u_{\mathrm{scr}}$ in (23) and the unscreened ones $\delta u_{\mathrm{unscr}}$ in (24) indicate that there exists a critical particle volume fraction $\phi_{\mathrm{crit}} \sim (L/a)^3 Re_p$ above which the backflow inertia screens velocity fluctuations under $\xi<L$ without being limited by the system size $L$. The transition from $\delta u_{\mathrm{unscr}} \propto \phi^{1/2}$ for $\phi < \phi_{\mathrm{crit}}$ to $\delta u_{\mathrm{scr}} \propto \phi^{1/3}$ for $\phi > \phi_{\mathrm{crit}}$ is qualitatively consistent with the sidewall-restraint picture proposed by Brenner [12]. The coexistence of screened and unscreened regimes also bears some phenomenological resemblance to the nonlinear stochastic description of Lavine *et al* [16].

These two distinct regimes can be continuously bridged [25] through the system size $L$ while keeping $\xi$ fixed at a given $Re_p$:

$$\frac{\delta u}{\delta u_{\mathrm{scr}}(\xi)} = \left[1+\left(\frac{\xi}{L}\right)^4\right]^{-1/8}. \qquad (25)$$

FIG.3 compares (25) with the data measured by Segré *et al.* [3]. The inset plots $\delta u/2\phi^{1/3}V_s$ against $L/a\phi^{-1/3}$, as originally displayed in their work. Using their plateau value $2\phi^{1/3}V_s$ for $\delta u_{\mathrm{unscr}}$ and crossover length $150a\phi^{-1/3}$ for $\xi$, (25) successfully captures the reported transition from $\delta u_{\mathrm{unscr}} \propto L^{1/2}$ for $L/\xi$<<1 to $\delta u_{\mathrm{scr}}(\xi) \propto L^0$ for $L/\xi$>>1.With $Re_p \approx 0.6\times10^{-4}$ in their study, the predicted plateau $\delta u_{\mathrm{scr}}\approx 10.3\phi^{1/3}V_s$ and crossover length $\xi \approx 46a\phi^{-1/3}$ are in good quantitative agreement with the measured values, differing only by $O(1)$ factors given the absence of adjustable parameters. In the main plot $\delta u/\delta u_{\mathrm{scr}}$ vs. $L/\xi$, we rescales the theoretical curve and experimental data by their own plateau value and crossover length, showing that both collapse onto the same universal curve described by (25).

A similar crossover and data collapse can also be obtained by varying $Re_p$ while keeping $L$ fixed:

$$\frac{\delta u}{\delta u_{\mathrm{unscr}}(L)} = \left[1+\left(\frac{Re_p}{Re_{p,\mathrm{TH}}}\right)^{4/3}\right]^{-1/8}. \qquad (26)$$

As shown by replotting the data of FIG.2 in FIG.4, (26) successfully reproduces the trend reported by Bergougnoux and Guazzelli [4]: the observed plateau $\delta u_{\mathrm{unscr}}$ for $Re_p<Re_{p,\mathrm{TH}}$ followed by $\delta u_{\mathrm{scr}} \propto Re_p^{-1/6}$ for $Re_p>Re_{p,\mathrm{TH}}$ (inset). With $\phi$=0.3% and $L/a\approx 260$ in their experiments, the theory predicts $\delta u_{\mathrm{unscr}} \approx 1.33V_s$ and $Re_{p,\mathrm{TH}}\approx 10^{-4}$, consistent in magnitude with the measured values 0.48 and $4\times10^{-4}$. The normalized plot of $\delta u/\delta u_{\mathrm{unscr}}$ vs. $Re_p/Re_{p,\mathrm{TH}}$ also collapses the theoretical and experimental curves into a single universal curve described by (26).

The successful collapse of both the $L$- and $Re_p$-dependent data onto their respective universal crossover curves shown above demonstrates that the backflow inertial screening mechanism correctly captures the universal transition between finite-system (unscreened) and finite-correlation (screened) regimes.

With the screened velocity fluctuations given by (23), the corresponding hydrodynamic self diffusivity can be readily obtained from the Green-Kubo relation with time integration up to the viscous correlation time $\tau_c=\xi^2/\nu=1/\omega^*$ from (16), as performed in the experiments by Nicolai and Guazzelli [7]. This gives $D_{H//}= 4\,D_{H\perp} = (\delta u)^2\,\tau_c$ of amplitude

$$D_H = \gamma S(0)\, V_s\, a, \tag{27}$$

which reproduces the well-known scaling observed in experiments [7,8], where $\gamma=\alpha^2 A'^{-2/3}=24\sqrt{3}/\pi\approx13.23$. For hard-sphere suspensions, the static structure factor in the compressibility limit is given by $S(0)=(1-\phi)^4/(1+2\phi)^2$ using the Percus-Yevick approximation. With $\phi$=0.05 in experiments [7,8], this yields $S(0)\approx0.87$ and $D_H\approx 11.57V_s a$, consistent in magnitude with the measured value $6V_s a$.

In conclusion, the proposed backflow inertia screening mechanism not only prevents the indefinite growth of velocity fluctuations through an emergent screening length, but also successfully accounts for experimental observations, with predictions consistent in magnitude with the measured values. The predicted velocity fluctuations and hydrodynamic self diffusivities are systematically larger than the corresponding experimental values. Such differences are not unexpected: several factors absent from the present theory may suppress velocity fluctuations and thereby reduce the numerical prefactors—most notably finite-boundary effects [26,27], spatial inhomogeneities arising from particle clustering and void regions [4], particle polydispersity [27-29], and density stratification [13-15]. Nevertheless, the predicted scaling laws, screening criterion, and crossover behavior remain robust, suggesting that the non-negligible inertia induced by the compensating backflow may play a central role in governing the spatiotemporal evolution of large-scale velocity fluctuations in sedimenting suspensions. These quantitative predictions may yield signatures testable in both experiments and simulations, inviting systematic, well-controlled investigations of how the particle Reynolds number and system size affect fluctuation correlations and crossover behavior [30].

This research is supported by National Science and Technology Council of Taiwan.

*hhwei@mail.ncku.edu.tw